\documentclass[prb,reprint, twocolumn]{revtex4-1} 
\usepackage{amsmath}
\usepackage{amsfonts} 
\usepackage{graphicx} 
\usepackage{ulem}
\usepackage{url}
\begin{document}

\title{Video recording true single-photon double-slit interference}

\author{Reuben S. Aspden}
\affiliation{SUPA, Department of Physics and Astronomy, University of Glasgow, Glasgow G12~8QQ, United~Kingdom}

\author{Miles J. Padgett}
\email{miles.padgett@glasgow.ac.uk}
\affiliation{SUPA, Department of Physics and Astronomy, University of Glasgow, Glasgow G12~8QQ, United~Kingdom}

\author{Gabriel C. Spalding}
\affiliation{Department of Physics, Illinois Wesleyan University, Bloomington, IL 61701, United~States}

\collaboration{QuantIC Collaboration}

\date{\today}


\begin{abstract}
As normally used, no commercially available camera has a low-enough dark noise to directly produce video recordings of double-slit interference at the photon-by-photon level, because readout noise significantly contaminates or overwhelms the signal. In this work, noise levels are significantly reduced by turning on the camera only when the presence of a photon has been heralded by the arrival, at an independent detector, of a time-correlated photon produced via parametric down-conversion. This triggering scheme provides the improvement required for direct video imaging of Young's double-slit experiment with single photons, allowing clarified versions of this foundational demonstration. Further, we introduce variations on this experiment aimed at promoting discussion of the role spatial coherence plays in such a measurement. We also emphasize complementary aspects of single-photon measurement, where imaging yields (transverse) position information, while diffraction yields the transverse momentum, and highlight the roles of transverse position and momentum correlations between down-converted photons, including examples of ``ghost'' imaging and diffraction. The videos can be accessed at \url{http://sun.iwu.edu/~gspaldin/SinglePhotonVideos.html} online.
\end{abstract}

\maketitle 

\section{Introduction}

To a great many, the word photon brings to mind a picture of a particle-like ball (or, perhaps, a ray that describes the ball's trajectory). Such a photon cannot exist. Yet these notions are so widespread that they have led to suggestions that physicists ought to receive special training and a license before being allowed to use the word ``photon.'' \cite{Lamb-1995} Such training would undoubtedly center upon discussion of Young's double slit experiment, which falls into a small class of experiments that have a special place in both the history and development of physics. Generically, illumination of a double slit with spatially coherent light (usually an expanded laser beam) creates sinusoidal fringes in the far field. These fringes persist even when the illumination light source is reduced in intensity such that no more than one photon is present within the region of the slits at any time, providing the archetypal example of single-particle interference and wave-particle duality. 

As enshrined in the uncertainty principle, duality is characterized by an intrinsic incompatibility of simultaneously defined position and momentum for a single particle. Whereas knowledge of \textit{which} slit the photon is transmitted through defines the photon's transverse position in the plane of the slits, knowledge of the photon's position in the far field (\textit{i.e.}, within the interference pattern) gives its transverse momentum. However, physically configuring the experiment in such a way as to provide knowledge of which slit the photon passes through \textit{prevents} observation of an interference pattern.\cite{Feynman-1965} The mainstream contemporary interpretation of quantum mechanics is that this incompatibility extends beyond considerations of measurement-induced perturbations, and is rooted in an inherent uncertainty associated with the particle itself.

In terms of the philosophical debate, rather than the ``which slit'' question it is the \textit{mechanism} by which the extended photon wavefunction is collapsed (or projected) into a specific location in any one of the many sinusoidal far-field fringes that is most contentious (the ``measurement problem of quantum mechanics''). Whereas Einstein raised\cite{Mermin-1985} key questions about the outcome of such an experiment, \textit{i.e.} whether or not ``hidden variables'' contained information about which fringe the photon would be observed in, the mainstream contemporary interpretation is that the outcome remains a probabilistic distribution until (at least) the moment of irreversible interaction with the detector.\cite{Griffiths-2005} Yet this notion that a single-photon event remains indefinitely in a superposition of different fringes until the moment of detection presents challenges to our conventional understanding. In the case to be presented here, using spatially separated detectors for an entangled \textit{pair} of photons, the nature of wavefunction collapse becomes even more interesting to consider, as the consequences of this collapse can be taken, incorrectly, to imply a  \textit{nonlocal} cause and effect!

\begin{figure*}[t]
\centering
\includegraphics[width = 6.5in]{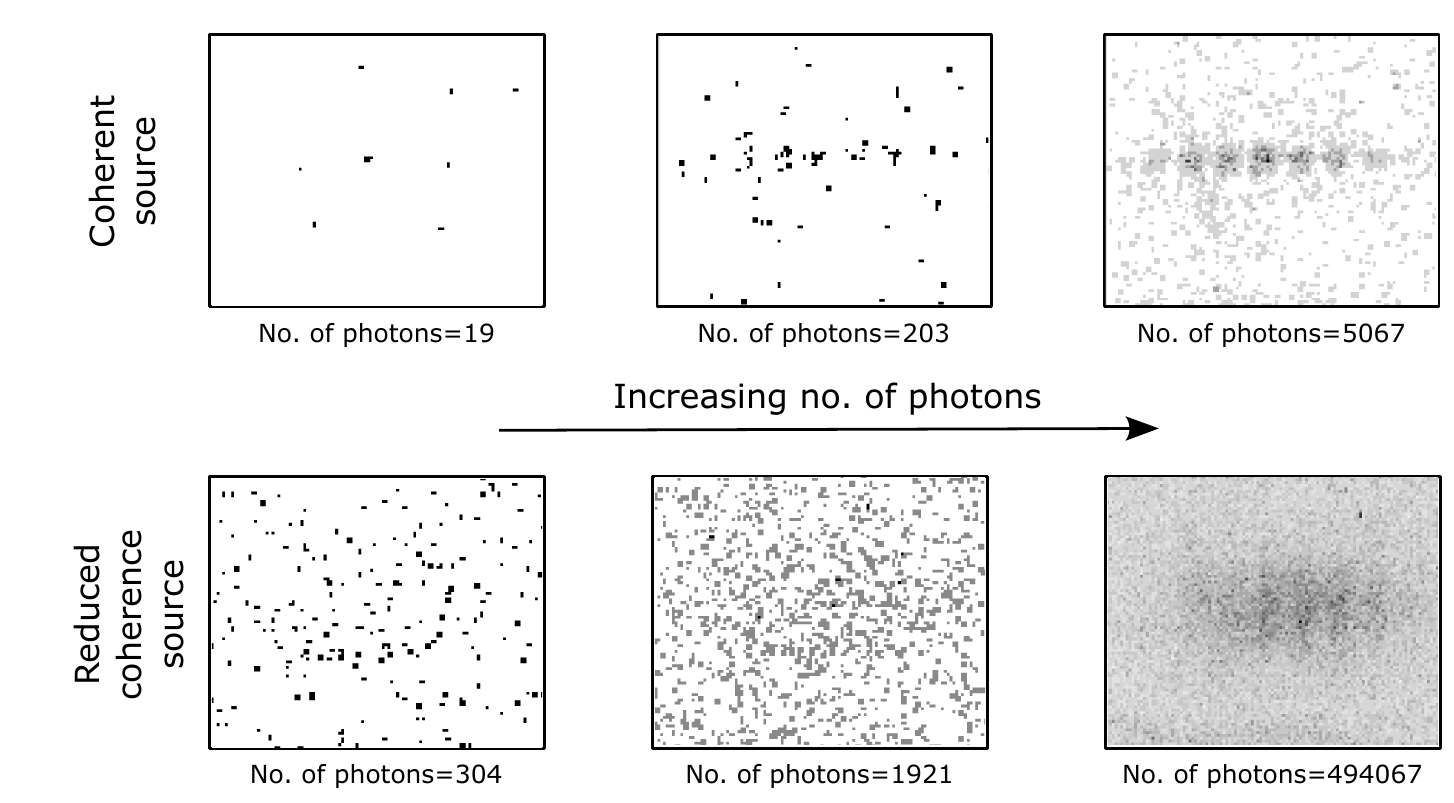}
\caption{Sample frames illustrating single-photon double-slit diffraction. The sequence along the top is for a case where spatial coherence has been experimentally enhanced, by only accepting photons associated with a single mode, as discussed in the text. For comparison, the multi-modal case is shown at bottom.}
\label{frames}
\end{figure*}

Despite the central conceptual role in modern physics that is played by single-photon interference, there have been few visualizations of actual data presented for Young's double-slit interference experiment. Indeed, none of these previous examples have included direct 2D video imaging of the distribution of single-photon detection events. The videos supplementing this article, which are intended for classroom use, aim to address this need, and can be found at \url{<http://sun.iwu.edu/~gspaldin/SinglePhotonVideos.html>} online. Representative frames are shown in Fig.~\ref{frames}.

It is essential that compelling experimental data be made available for the purpose of introducing students to foundational phenomena, especially when those phenomena differ markedly from expectations based upon common experience, as is often the case with quantum physics. Quantum Physics is one of the conceptual pillars of the physics curriculum, and is the required framework for understanding (and teaching) a broad range of topics. For example, Intel has released chipsets based upon 14~nm components, and is in pre-production of 10~nm-linewidth designs; at such scales, classical rules of conduction are superseded, and so any student thinking of becoming an electrical engineer should be introduced, at some level, to quantum principles. Similarly, teaching Materials Physics often requires establishing a foundational understanding of basic quantum mechanics. Students are fascinated by the potential for quantum technologies to enable secure transmission of data (quantum encryption), and to revolutionize computing architectures (quantum computing). Connections to such areas of opportunity can motivate student engagement, but abstract formalism and simulation are not, alone, sufficient: pedagogical discussions about experimental observations can help to assure students that the dialog is well grounded.

\section{Correlated Photons}

\begin{figure} [h]
\centering
\includegraphics[width = 3in]{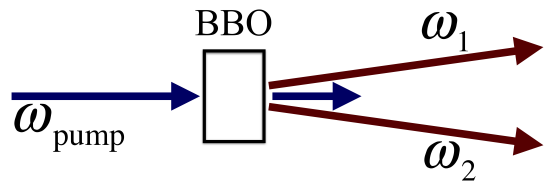}
\caption{Down-conversion: $\hbar\omega_{pump}=\hbar\omega_{1}+\hbar\omega_2$. A single 355~nm photon can be absorbed and re-emitted as a pair of quantum mechanically \textit{entangled} photons, each with wavelength 710~nm, chosen because it is the wavelength of peak quantum efficiency for the detectors used. However, even if pump beam is well collimated, the output is not.}
\label{downconv}
\end{figure}

In this paper, critical use is made of pairs of photons that are produced by parametric down-conversion within $\beta$-barium borate crystals (hereafter ``BBO''). There are a number of introductions to downconversion,\cite{SPDC-intro1, SPDC-intro2} but here we highlight several constraints placed upon the down-conversion process. First, down-conversion replaces one photon with two new photons. The energies of the emitted pair of photons must, according conservation of energy, add up to the energy of the incident pump photon: $\hbar\omega_{pump}=\hbar\omega_{1}+\hbar\omega_2$. This implies that the pair of photons must have been emitted from the same spatial position within the crystal where the first photon is absorbed. Moreover, the momenta of the emitted pair of photons must add up to the momentum of the incident photon. These constraints upon position and momentum might seem to challenge the uncertainty principle which, as noted above, states that position and momentum cannot be simultaneously well-defined for any single photon. 

However, because the conservation principles only constrain the summed properties of the two down-converted photons, they cannot be described individually, but only as an ``entangled'' two-particle wavefunction. For example, though the fact that the \textit{sum} of their momenta is fixed means that their transverse momenta must add to zero, the momentum of each individual photon is not defined; the two-particle wavefunction itself contains a \textit{superposition} of individual momentum values for each photon. In a similar fashion, properties such as the energy, angular momentum, and polarization of each individual photon in the pair can be engineered, depending upon the experimental design, to \textit{have no definite values} until a measurement is made.\cite{Zurek-2007,Zurek-1991} (Entanglement persists over the length scales probed by our experiments.\cite{Tittel-1999,EndNote-1})

While the two well-defined rays sketched in Fig.~\ref{downconv} might suggest that the down-converted photons should emerge at well-defined angles, experiments show that light emitted from the down-conversion process always has an angular range. In our experiment, the light emitted at $\omega_{1}$ and the light emitted at $\omega_{2}$ are essentially \textit{collinear}, but there remains a spread in the angular distribution that arises due to geometric effects associated with the length of the down-conversion crystal.\cite{Glasgow-2012} (For the experiments shown here, we used a 3~mm thick BBO crystal.) Because the light spreads more quickly than would be the case for a Gaussian mode, the BBO crystal must be considered a multi-modal source. However, as demonstrated below, production of an interference pattern with high-contrast fringes requires the high degree of spatial coherence associated with a single-mode source. 

\section{Experimental Setup}

\begin{figure}[h]
\centering
\includegraphics[width = 3.55in]{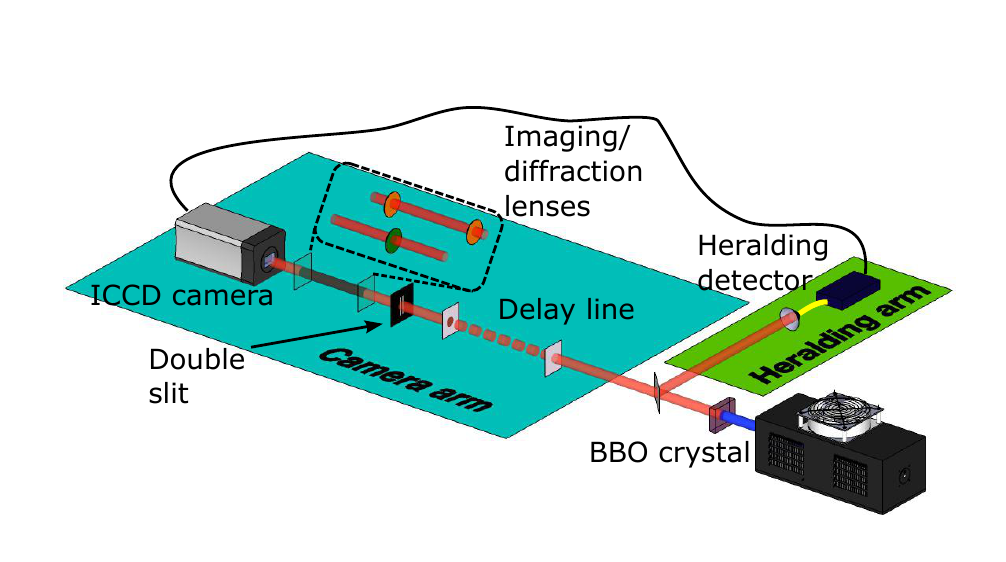}
\caption{Simplified schematic. Following a beamsplitter, two collection arms can detect the correlated photons emerging from the BBO crystal. The signal-to-noise ratio is improved by using the heralding detector to only trigger the time-gated intensified CCD (ICCD) camera when a time-correlated photon is due to arrive. The two slits used here were each 100~microns wide, with a center-to-center separation of 500~microns.}
\label{Schem1}
\end{figure}

Fig.~\ref{Schem1}, illustrating our basic experimental approach, features a pump laser at far right, followed by the BBO crystal, from which we extract down-converted photon pairs (and block any transmitted light at the pump frequency). The time-gated, intensified CCD (ICCD) camera at far left is capable of detecting the down-converted photons but, to reduce noise, needs to be time gated: that's the purpose of the beamsplitter and the heralding arm. The camera will only record data in those instances where one photon takes the path to the heralding detector, which provides the trigger signal to the camera's gate, while the other member of the down-converted pair goes on to be detected by the camera. Along the way to the camera, this photon encounters the slits. In order to produce a high-contrast interference pattern, there needs to be some workaround for the fact that the BBO crystal is a multi-modal source, and so lacks the required degree of spatial coherence. The key is to use a single-mode optical fiber to limit what light makes it into the heralding detector, and exploit the fact that the camera will only detect down-converted photons that are highly correlated with those detected by the heralding detector: essentially, it is \textit{as if} the slits were being illuminated by a single-mode source.

Some further details of this experimental apparatus may be of interest. In our experiment we use type-I spontaneous parametric downconversion (SPDC) in a \textit{collinear} regime, which is why we require a beamsplitter to separate the output photons. This separation occurs in a probabilistic manner: separation efficiency at the beamsplitter is only 50\%. However, as this loss is before the photons interact with the object it does not affect the heralding efficiency of the system. The down-converted photon pairs are separated from the remaining pump beam after the BBO crystal by use of a ``cold mirror,'' which is the term used for a standard, commercially available dielectric mirror that reflects short wavelengths while very efficiently transmitting infrared wavelengths. The reflected pump beam was sent to a beam dump, while the transmitted light was filtered by a 10~nm bandwidth high transmission interference filter centered on 710~nm to ensure that only the down-converted photons are present in our system. The slits used here were 100~microns wide, with a center-to-center separation of 500~microns. The heralding detector we have used is a fiber-coupled single-pixel, single-photon avalanche diode (``SPAD'') with quantum efficiency at 710~nm of approximately 65\%. Detection of a photon produces a short (15~ns) TTL pulse that is used to trigger the camera. An optical delay line added to the camera arm compensates for the electronic delay associated with this triggering. Measured dark-count rates for this SPAD were roughly 100 per second, but these would normally only result in a blank image at the camera. Crucially, the gating time on the camera is limited to a 5~ns coincidence window, so the contribution of noise from the camera photocathode is also minimal. The photon flux used for this experiment was such that the camera detects only the single, correlated photon in each coincidence window. The signals from multiple coincidence windows can be accumulated on the CCD chip over a longer exposure period before the frame is read out. Individual frames may therefore contain multiple photon events, although the accumulation time was always such that there was statistically much less than one photon per pixel in any frame. For presentation purposes, during post-processing the videos have been recompiled such that the photon arrival rate increases exponentially as the video progresses. 

Except for the camera, which is an (Andor iStar gen III) intensified CCD, hundreds of institutions already have all of the elements of the set-up shown, in undergraduate instructional labs. For example, educationally priced SPADs are available from the Advanced Laboratory Physics Association (ALPhA).\cite{ALPhAsite} Regardless, any institution may make use of the video data provided with this article. 

As possible extensions of this work, we note that it is possible to perform similar studies, with compromised spatial resolution and loss of 2D information, by using multiple SPADs arrayed along a line rather than a two-dimensional ICCD camera.\cite{Copernicus-2014} Such experiments could be performed even more affordably by using a single, scanning-fiber detector in place of the ICCD camera. However, when using a scanning detector, each image would have to be built up from exposures taken with the detector in \textit{N} different locations, yielding \textit{N} distinct pixels in the reconstructed image. The detection efficiency of the imaging system would fundamentally be limited to a maximum of 1/\textit{N}, thus increasing the collection time by a factor of \textit{N}. Here, this limitation is overcome by replacing the scanning detector by a two-dimensional ICCD camera, thereby detecting all photons irrespective of their position within the image, an approach which opens up many opportunities.\cite{Woerdman-2002,Zeilinger-2013} All of this begs the question: what is the minimum number of photons required to form an image? With conventional cameras, of order $10^{12}$ photons would be collected for each image.\cite{Llull-2013} For a megapixel camera, this corresponds to $10^{6}$ photons per pixel. Our paper describes two distinct imaging modes, each utilizing the same components: heralded imaging and ghost imaging.\cite{Shapiro-Boyd-2012} In both configurations it is possible to produce high-quality images from an average of much fewer than one detected photon per image pixel.\cite{Morris-2015} 

\section{Heralded measurements}

In our heralded imaging system, we exploit the strong temporal correlations between the down-converted photons to obtain photon-by-photon measurements of an image of a double slit and, in a separate experiment, we capture the diffraction pattern obtained from the single photons passing through the slits. Again, without such heralding, the readout noise associated with any commercially available camera would significantly corrupt or completely swamp the single-photon signal of interest.\cite{Harvard-2013,Walmsley-2009} 

The optics following the BBO crystal re-images the down-converted light onto the double slits. So, when the camera is located in an image plane of the BBO crystal (as is the case when using the optics represented by the lenses shown in orange in Fig.~\ref{Schem1}) we capture an image, one photon at time, of the slits, rather than the interference pattern produced by those slits. Conversely, by substituting the 500~mm lens shown in green in Fig.~\ref{Schem1} for the two 250~mm lens shown in orange, we can instead capture the double-slit interference pattern (also one photon at a time).  

\begin{figure}
\centering
\includegraphics[width = 3.5in]{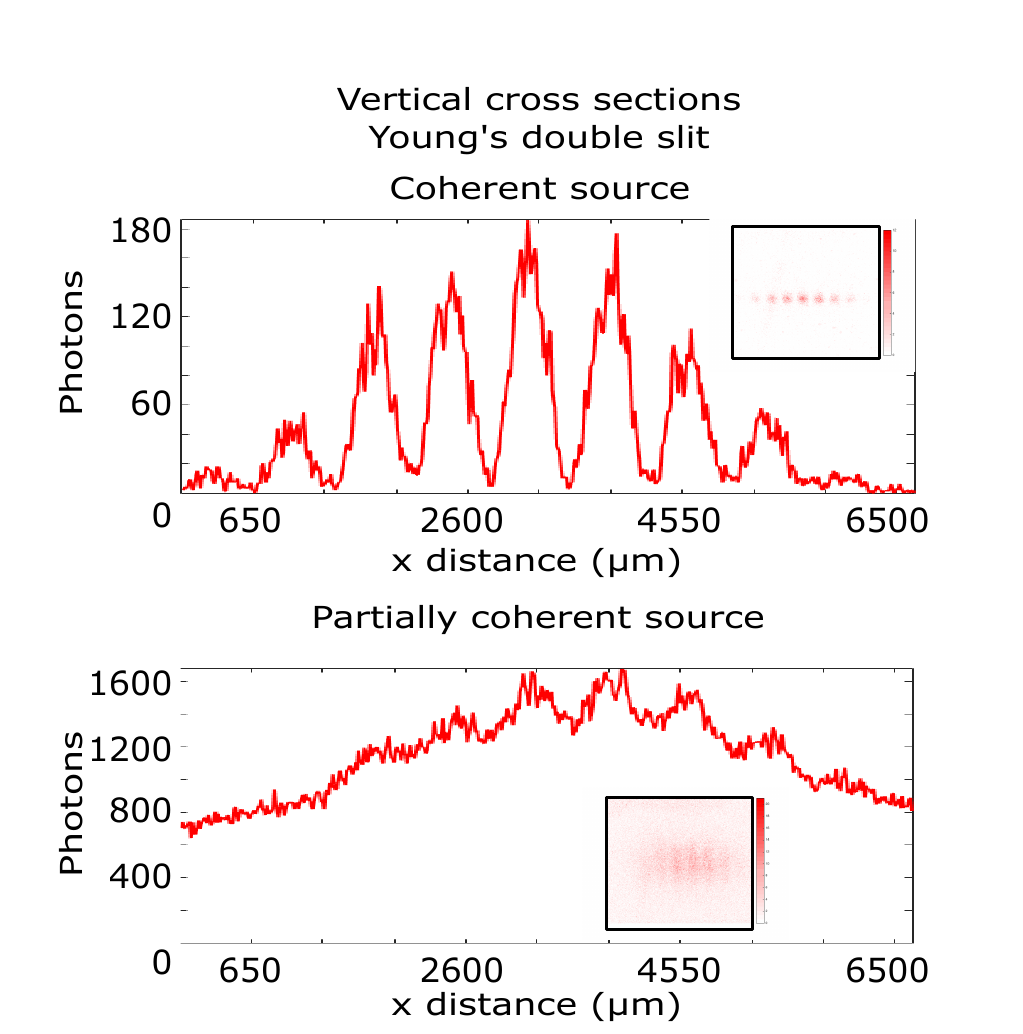}
\caption{Top: Diffraction pattern for a case where spatial coherence has been experimentally enhanced, by only accepting photons associated with a single mode. For comparison, the multi-modal case is shown at bottom. (Insets show the full 2D datasets.)}
\label{sum_diff}
\end{figure}

\begin{figure}
\centering
\includegraphics[width = 3.5in]{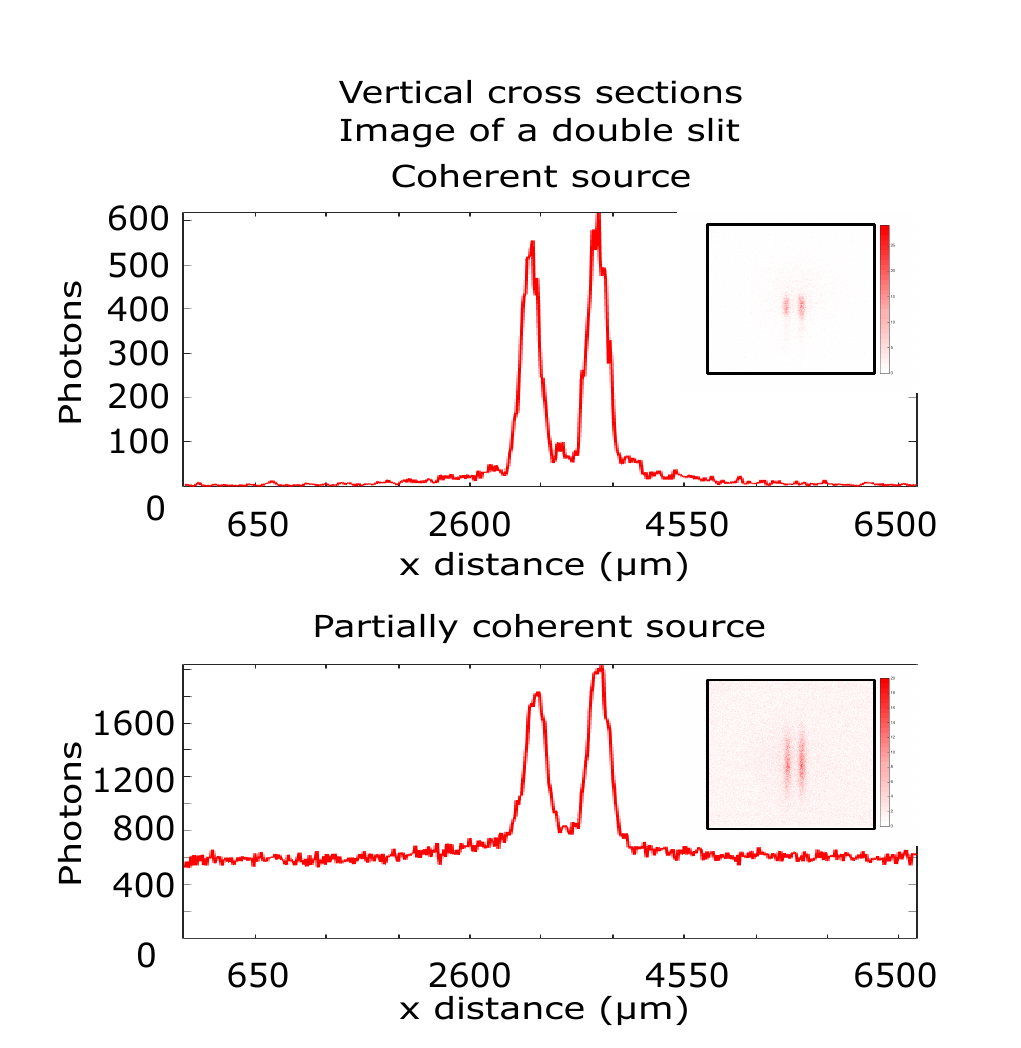}
\caption{Top: Image results when using single-mode fiber. Bottom: comparable results obtained from multi-mode fiber.}
\label{sum_im}
\end{figure}

To promote discussion of the role of spatial coherence, we show the effects of replacing the single-mode optical fiber that couples light into our heralding detector with multi-mode fiber. Accepting this broader range of heralded photons means that the slits themselves are effectively being illuminated with a multi-mode source. As a consequence the diffraction patterns are largely erased, as shown in Fig.~\ref{sum_diff}, while imaging is far less affected, as demonstrated in Fig.~\ref{sum_im} (though the increased vertical spread of the detected signal should be noted, as it is indicative of the multi-modal illumination).

Simply put, imaging tells us about the transverse (horizontal) spatial location of photons within the plane of the object being imaged. Because the location of the slits in that object are the same, regardless of whether we use single-mode or multi-mode fiber, the imaging results can be roughly comparable. On the other hand, diffraction depends upon the transverse momenta of the photons as they exit those slits. Clearly, the diffraction pattern is sensitive to the degree of spatial coherence, while imaging is robust.

\section{Spatially correlated ghost measurements}

\begin{figure}[h]
\centering
\includegraphics[width = 3.55in]{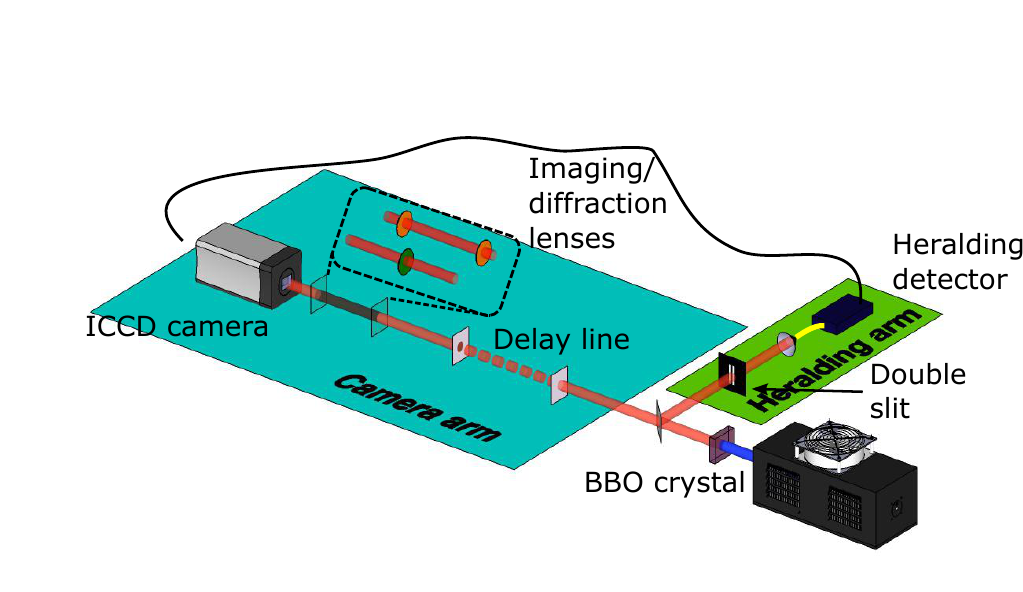}
\caption{Ghost imaging schematic. The slits are no longer in the same arm as the camera, yet when data collected by the camera is triggered only by arrival of a correlated photon at the heralding detector, the results contain information about the slits.}
\label{GSchem}
\end{figure}

Having discussed the importance of temporal correlations, we turn to another configuration, shown in Fig.~\ref{GSchem}, designed to highlight the importance of nonlocal spatial correlations. Specifically, we also took data when our slits were placed into the arm that does \textit{not} contain the camera. Using this version of the set up, even though the photons collected by the camera \textit{have never interacted with} the two slits, it is nevertheless possible to perform experiments revealing information about either imaging or diffraction. Such experiments are referred to as ghost imaging and ghost interference,\cite{Uni.Genova-2007} with the word ``ghost'' referencing Einstein's concern over ``spooky action at a distance.''\cite{Mermin-1985}

\begin{figure}[h]
\centering
\includegraphics[width = 3.5in]{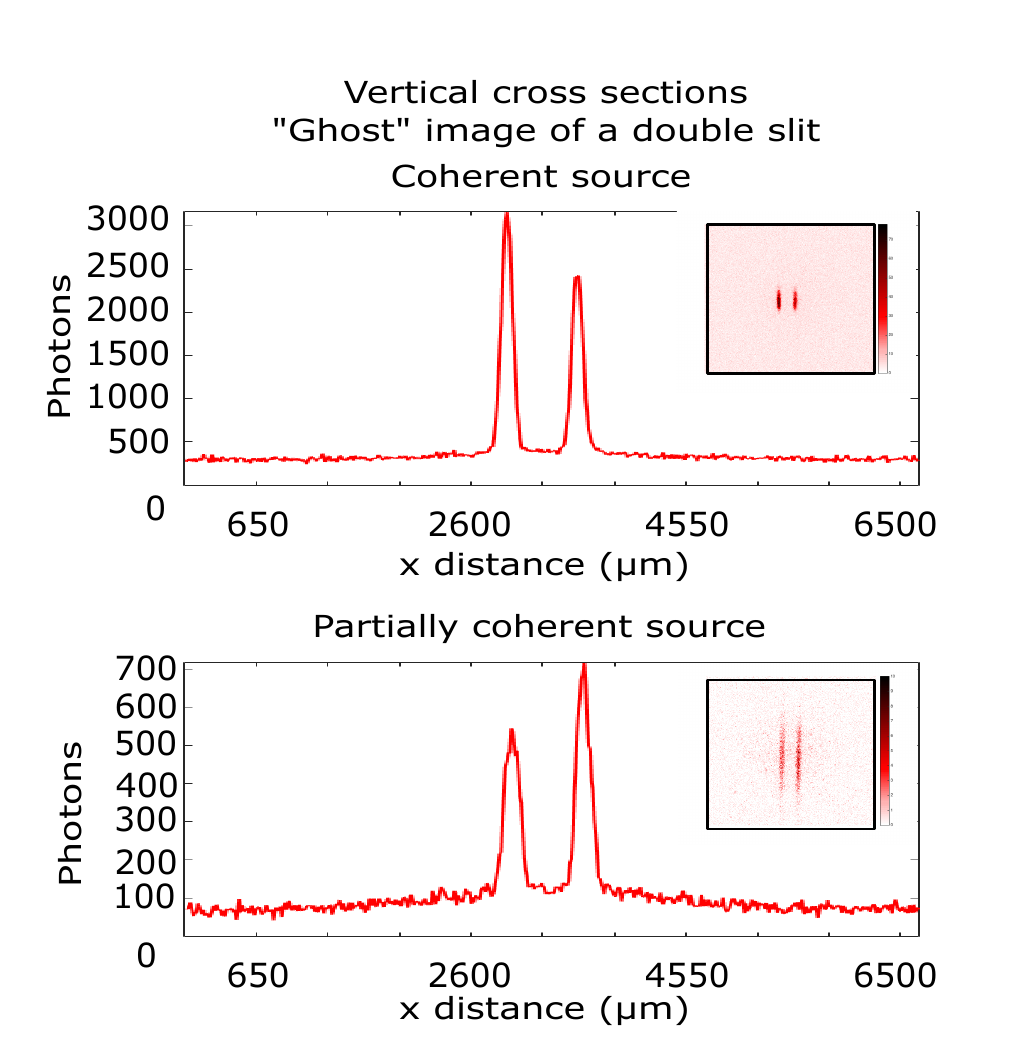}
\caption{Ghost imaging data. According to any classical model, the photons used to construct this image have never interacted with the object imaged. An image is formed is due to the spatial correlation between entangled photons.}
\label{GIm}
\end{figure}

For ghost imaging, illustrated Fig.~\ref{GIm}, the key point is that the down-converted photons pairs are not just temporally correlated, but also spatially correlated. Here, the plane containing the slits in the heralding beam and the plane of the ghost image are conjugate image planes. Yet, were it not for the heralded triggering employed, the complete set of photons hitting the ICCD camera would reveal only the incident (Gaussian) beam shape. Nor could the data collected by the SPAD (which contains no spatial information), taken on its own, reveal an image. Rather, it is the correlation between the two data sets that allows us to select out those photons required for image construction.

\begin{figure}[h]
\centering
\includegraphics[width = 3.5in]{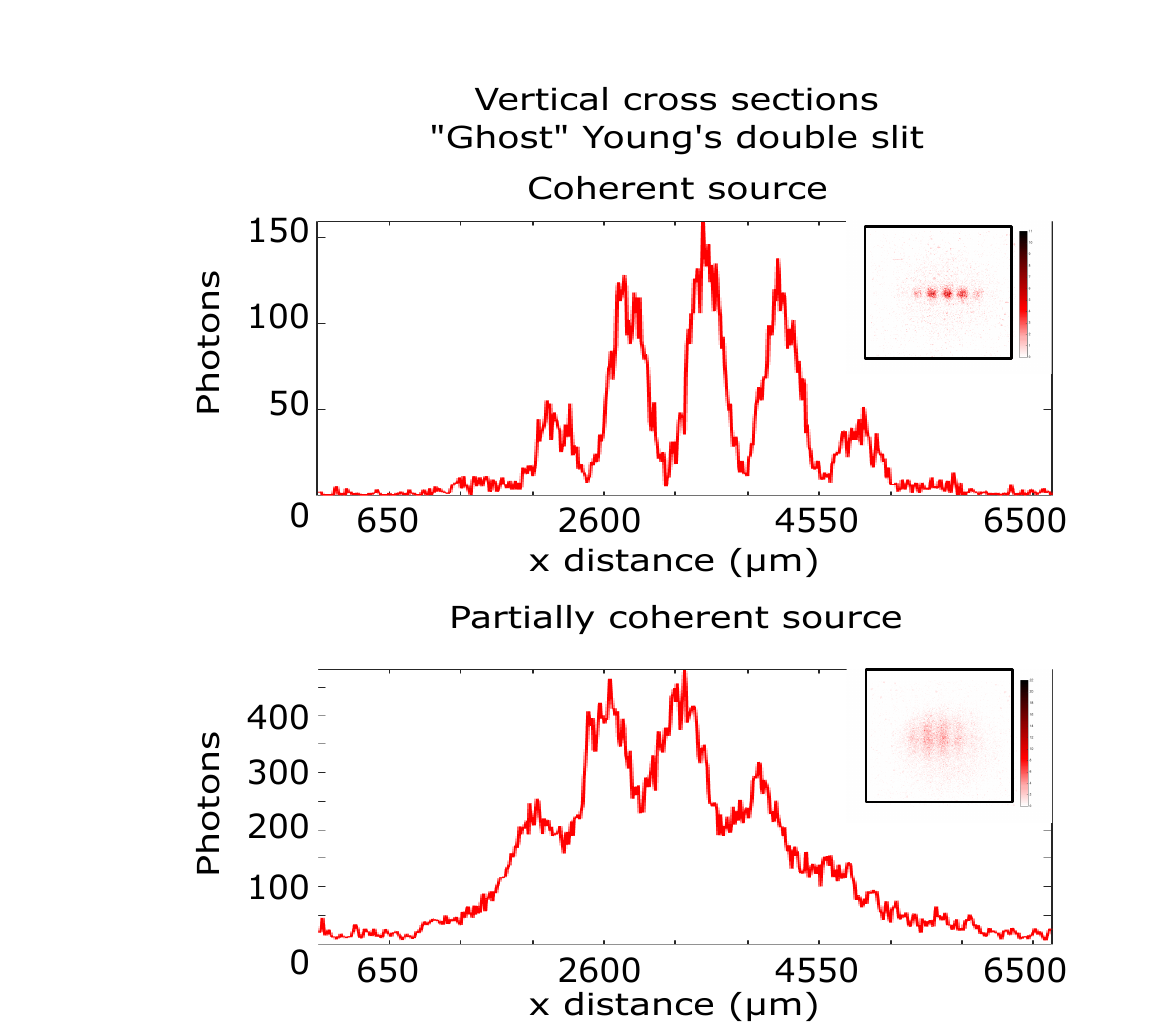}
\caption{Ghost diffraction data collected photon by photon.}
\label{GDiff}
\end{figure}

Opportunities for discussion are extended much further when we reconfigure our setup so as to examine the far field, where we expect (according to the classical schematic in Fig.~\ref{downconv}) the photons to have diverged. By inserting the optics represented by the lens shown in green in Fig.~\ref{GSchem}, we capture the ghost far-field double-slit interference pattern as shown in Fig.~\ref{GDiff}, rather than a direct image of the object containing the slits. 

The spacing of the fringes in a diffraction pattern can be calculated using the slit geometry and the distance from the diffracting object to the detector. For the case of ghost diffraction, it may not be immediately obvious what distance to use. An appeal to the literature shows that, in order to fit the fringe spacing observed in ghost diffraction, earlier experimental configurations made use of the distance from the diffracting object back through the beam splitter to the (finite extent) BBO crystal and then along the path to the ICCD camera (including any intervening lenses), \textit{as if} the light were traveling backwards in time.\cite{Klyshko-Shih-1995} This ``Klyshko model'' has found broad applicability to ghost optics.\cite{Glasgow-2014} However, it would be incorrect to conclude that the collapse of the two-particle wavefunction implies \textit{nonlocal} cause and effect. Ghost diffraction illustrates how the uncertainty principle applies to the two-photon wavefunction (rather than to the two photons individually): a constraint placed upon transverse position within the heralding arm reveals transverse momentum information in the ghost arm. This may seem a bit more spooky than the position correlations revealed in earlier portions of this paper, but is the full extent of the nonlocal weirdness. In our experiment, the signal distribution in the ghost diffraction plane is simply the Fourier transform of the amplitude of the distribution found in the ghost image plane. Thus, the fringe spacing observed is determined by the effective focal length of the optics utilized. Moreover, given the optical layout incorporated in the experiments presented here, the ghost diffraction plane and the heralded diffraction plane are 1:1 equivalent planes. That is, in our observations the fringe spacing in ghost diffraction are \textit{the same} as those observed in heralded diffraction. Students should be encouraged to compare the observed fringe spacing to the theoretical prediction, which can be calculated to be $f \lambda/d$, where \textit{d} is the center-to-center distance between the slits, $\lambda$ is the wavelength of the down-converted light, and \textit{f} is the 500~mm focal length of the lens which lies between the image plane of the slits and the far-field interference pattern. Again, the double slit used here had a center-to-center separation of 500~microns.

\section{Conclusion}


Quantum mechanics is full of counter-intuitive phenomena. For this reason, we believe it to be useful that real experimental data be used to introduce students to some of the foundational phenomena of quantum mechanics. The processed video versions of the data described in this article are available at \url{<http://sun.iwu.edu/~gspaldin/SinglePhotonVideos.html>} online, and are intended for incorporation into classroom discussions. In support of those discussions, we highlight some of the more accessible articles that might be of use to this audience. Down-converted photons play an important role in a series of instructional labs in quantum mechanics that are becoming a widespread and important part of the undergraduate curriculum in physics\cite{Colgate-2005, DickinsonCollege-2010, Colgate-2014, Beck-2012} (providing evidence for the existence of photons\cite{WhitmanCollege-2004,Colgate-2005} and the tendency of bosons (such as photons) to bunch together,\cite{WhitmanCollege-2004} as well as addressing, \textit{e.g.}, how one might produce entangled photons for quantum computing and teleportation,\cite{Prutchi-2012} or for disproving local realism,\cite{ReedCollege-2002a, ReedCollege-2002b, WhitmanCollege-2006} \textit{etc.}\cite{HOM-2012,Portland-2016}).

\begin{acknowledgments}

This work was funded by the UK EPSRC through a Program Grant (EP/M01326X/1). Raw data can be found in open-access repository at DOI \url{<http://dx.doi.org/10.5525/gla.researchdata.281>}. We wish to disclose that G.C.S. is on the Board of Directors for ALPhA, a non-profit organization, coordinating their program which makes educationally priced SPADs available.\cite{ALPhAsite}

\end{acknowledgments}


\begin{thebibliography}{0}%
\makeatletter
\providecommand \@ifxundefined [1]{%
 \@ifx{#1\undefined}
}%
\providecommand \@ifnum [1]{%
 \ifnum #1\expandafter \@firstoftwo
 \else \expandafter \@secondoftwo
 \fi
}%
\providecommand \@ifx [1]{%
 \ifx #1\expandafter \@firstoftwo
 \else \expandafter \@secondoftwo
 \fi
}%
\providecommand \natexlab [1]{#1}%
\providecommand \enquote  [1]{``#1''}%
\providecommand \bibnamefont  [1]{#1}%
\providecommand \bibfnamefont [1]{#1}%
\providecommand \citenamefont [1]{#1}%
\providecommand \href@noop [0]{\@secondoftwo}%
\providecommand \href [0]{\begingroup \@sanitize@url \@href}%
\providecommand \@href[1]{\@@startlink{#1}\@@href}%
\providecommand \@@href[1]{\endgroup#1\@@endlink}%
\providecommand \@sanitize@url [0]{\catcode `\\12\catcode `\$12\catcode
  `\&12\catcode `\#12\catcode `\^12\catcode `\_12\catcode `\%12\relax}%
\providecommand \@@startlink[1]{}%
\providecommand \@@endlink[0]{}%
\providecommand \url  [0]{\begingroup\@sanitize@url \@url }%
\providecommand \@url [1]{\endgroup\@href {#1}{\urlprefix }}%
\providecommand \urlprefix  [0]{URL }%
\providecommand \Eprint [0]{\href }%
\providecommand \doibase [0]{http://dx.doi.org/}%
\providecommand \selectlanguage [0]{\@gobble}%
\providecommand \bibinfo  [0]{\@secondoftwo}%
\providecommand \bibfield  [0]{\@secondoftwo}%
\providecommand \translation [1]{[#1]}%
\providecommand \BibitemOpen [0]{}%
\providecommand \bibitemStop [0]{}%
\providecommand \bibitemNoStop [0]{.\EOS\space}%
\providecommand \EOS [0]{\spacefactor3000\relax}%
\providecommand \BibitemShut  [1]{\csname bibitem#1\endcsname}%
\let\auto@bib@innerbib\@empty
\end{thebibliography}%


\begin{thebibliography}{99}

\bibitem{Lamb-1995} W. E. Lamb, Jr., ``Anti-photon,'' Appl. Phys. B \textbf{60}, 77--84 (1995).

\bibitem{Feynman-1965} R. P. Feynman, \textit{The Character of Physical Law - part 6 probability and uncertainty}, a series of lectures recorded by the BBC at Cornell University. \url{<https://youtu.be/aAgcqgDc-YM>}

\bibitem{Mermin-1985} N. David Mermin, ``Is the moon there when nobody looks? Reality and the quantum theory,'' Physics Today \textbf{38}, 38--47 (1985).

\bibitem{Griffiths-2005} David J. Griffiths, \textit{Introduction to Quantum Mechanics}, 2nd edition (Prentice Hall, Englewood Cliffs, NJ, 2005), pp. 2--4.

\bibitem{SPDC-intro1} ``Spontaneous parametric down-conversion,'' Wikipedia \url{<https://en.m.wikipedia.org/wiki/Spontaneous_parametric_down-conversion>}.

\bibitem{SPDC-intro2} J. Schneeloch, J. C. Howell, ``Introduction to the Transverse Spatial Correlations in Spontaneous Parametric Down-Conversion through the Biphoton Birth Zone,'' \url{<http://arxiv.org/abs/1502.06996>}.

\bibitem{Zurek-2007} W. H. Zurek, ``Decoherence and the transition from quantum to classical -- revisited,'' in \textit{Progress in Mathematical Physics}, \textbf{48}, edited by B. Duplantier, J. M. Raimond, and V. Rivasseau (Birkhauser Boston, Cambridge, MA, 2007), 1--31.

\bibitem{Zurek-1991} W. H. Zurek, ``Decoherence and the transition from quantum to classical,'' Phys. Today \textbf{44}, 36--44 (1991).

\bibitem{Tittel-1999} W. Tittel, J. Brendel, N. Gisin, H. Zbinden, ``Long-distance Bell-type tests using energy-time entangled photons,'' Phys. Rev. A \textbf{59}, 41504163 (1999).

\bibitem{EndNote-1} It is because of their weak interaction with their environments (in transparent media, at least) that photon-based experiments are preferred over electron transport studies for pedagogical labs. In fact, because BBO is highly transparent down-conversion will only occur a very tiny fraction of the time at the light intensities used in our experiments. Roughly one in $10^{11}$ photons gets ``down-converted,'' corresponding to a very small nonlinear optical susceptibility (\textit{i.e.}, the $\chi{_2}$ coefficient) even for a highly polarizable crystal such as BBO. (For this material $\chi{_3}$ is negligible.)

\bibitem{Glasgow-2012} F. M. Miatto, D. Giovannini, J. Romero, S. Franke-Arnold, S. M. Barnett, M. J. Padgett, ``Bounds and optimisation of orbital angular momentum bandwidths within parametric down-conversion systems,'' Eur. Phys. J. D \textbf{66}, 178 (2012).

\bibitem{ALPhAsite} Advanced Laboratory Physics Association (ALPhA) Web Site for dissemination of physics instructional lab materials beyond the first year of university:  \url{<http://www.advlab.org/spqm.html>}.

\bibitem{Copernicus-2014} P. Kolenderski, C. Scarcella, K. D. Johnsen, D. R. Hamel, C. Holloway, L. K. Shalm, S. Tisa, A. Tosi, K. J. Resch, T. Jennewein, ``Time-resolved double-slit interference pattern measurement with entangled photons,'' Scientific Reports \textbf{4}, 4685 (2014).

\bibitem{Woerdman-2002} S. S. R. Oemrawsingh, W. J. van Drunen, E. R. Eliel, J. P. Woerdman, ``Two-dimensional wave-vector correlations in spontaneous parametric downconversion explored with an intensified CCD camera,'' J. Opt. Soc. Am. B \textbf{19}, 2391--2395 (2002).

\bibitem{Zeilinger-2013} R. Fickler, M. Krenn, R. Lapkiewicz, S. Ramelow, A. Zeilinger, ``Real-time imaging of quantum entanglement,'' Scientific Reports \textbf{3}, 1914 (2013).

\bibitem{Llull-2013} P. Llull, X. Liao, X. Yuan, J. Yang, D. Kittle, L. Carin, G. Sapiro, D. J. Brady, ``Coded aperture compressive temporal imaging,'' Opt. Express \textbf{21}, 10526--10545 (2013).

\bibitem{Shapiro-Boyd-2012} J. H. Shapiro, R. W. Boyd, ``The physics of ghost imaging,'' Quantum Inf. Process. \textbf{11}, 949--993 (2012).

\bibitem{Morris-2015} P. A. Morris, R. S. Aspden, J.E.C. Bell, R. W. Boyd, M. J. Padgett, ``Imaging with a small number of photons,'' Nature Comm. \textbf{6}, 5913 (2015).

\bibitem{Harvard-2013} W. Rueckner, J. Peidle, ``Young's double-slit experiment with single photons and quantum eraser,'' Am. J. Phys. \textbf{81}, 951--958 (2013).

\bibitem{Walmsley-2009} L. Zhang, L. Neves, J. S. Lundeen, I. A. Walmsley, ``A characterization of the single-photon sensitivity of an electron multiplying charge-coupled device,'' J. Phys. B: At. Mol. Opt. Phys. \textbf{42}, 114011 (2009).

\bibitem{Uni.Genova-2007} L. Basano, P. Ottonello, ``Ghost imaging: open secrets and puzzles for undergraduates,'' Am. J. Phys. \textbf{75}, 343--351 (2007).

\bibitem{Klyshko-Shih-1995} D. V. Strekalov, A. V. Sergienko, D. N. Klyshko, Y. H. Shih, ``Observation of two-photon 'ghost' interference and diffraction,'' Phys. Rev. Lett. \textbf{74}, 3600--3603 (1995).

\bibitem{Glasgow-2014} R. S. Aspden, D. S. Tasca, A. Forbes, R. W. Boyd, M. J. Padgett, ``Experimental demonstration of Klyshko’s advanced-wave picture using a coincidence-count based, camera-enabled imaging system,'' J. Mod. Opt. \textbf{61}, 547--551 (2014).

\bibitem{Colgate-2005} E. J. Galvez, C. H. Holbrow, M. J. Pysher, J. W. Martin, N. Courtemanche, L. Heilig, J. Spencer, ``Interference with correlated photons: Five quantum mechanics experiments for undergraduates,'' Am. J. Phys. \textbf{73}, 127--140 (2005).

\bibitem{DickinsonCollege-2010} B. J. Pearson, D. P. Jackson, ``A hands-on introduction to single photons and quantum mechanics for undergraduates,'' Am. J. Phys. \textbf{78}, 471--484 (2010).

\bibitem{Colgate-2014} E. J. Galvez, ``Resource Letter SPE-1: single-photon experiments in the undergraduate laboratory,'' Am. J. Phys. \textbf{82}, 1018--1028 (2014).

\bibitem{Beck-2012} M. Beck, \textit{Quantum Mechanics: Theory and Experiment} (Oxford University Press, Oxford, 2012).

\bibitem{WhitmanCollege-2004} J. J. Thorn, M. S. Neel, V. W. Donato, G. S. Bergreen, R. E. Davies, M. Beck, ``Observing the quantum behavior of light in an undergraduate laboratory,'' Am. J. Phys. \textbf{72}, 1210--1219 (2004).

\bibitem{Prutchi-2012} D. Prutchi, S. R. Prutchi, \textit{Exploring quantum physics through hands-on projects} (Wiley, 2012).

\bibitem{ReedCollege-2002a} D. Dehlinger, M.W. Mitchell, ``Entangled photon apparatus for the undergraduate laboratory,'' Am. J. Phys. \textbf{70}, 898--902 (2002).

\bibitem{ReedCollege-2002b} D. Dehlinger, M.W. Mitchell, ``Entangled photons, nonlocality, and Bell inequalities in the undergraduate laboratory,'' Am. J. Phys. \textbf{70}, 903--910 (2002).

\bibitem{WhitmanCollege-2006} J. A. Carlson, M. D. Olmstead, M. Beck, ``Quantum mysteries tested: An experiment implementing Hardy's test of local realism,'' Am. J. Phys. \textbf{74}, 180--186 (2006).

\bibitem{HOM-2012} J. Carvioto-Lagos, G. Armendariz P, V. Velazquez, E. Lopez-Moreno, M. Grether, E. J. Galvez, ``The Hong-Ou-Mandel interferometer in the undergraduate laboratory,'' Eur. J. Phys. \textbf{33}, 1843--1850 (2012).

\bibitem{Portland-2016} J. M. Ashby, P. D. Schwarz, M. Schlosshauer, ``Delayed-choice quantum eraser for the undergraduate laboratory,'' Am. J. Phys. 84 \textbf{33}, 95--105 (2016).

\end{thebibliography}
\end{document}